\documentclass[prb,preprint,superscriptaddress]{revtex4-1}
\usepackage{times}
\usepackage{bm,graphicx,graphics,amsmath,amssymb,bm,epsfig,color}
\usepackage{euscript,tabularx}
\usepackage{textcomp}
\usepackage{longtable}

\begin{document}

\title{Optimal control of vortex core polarity by resonant microwave
  pulses}

\author{Benjamin Pigeau}
\affiliation{Service de Physique de l'\'Etat Condens\'e (CNRS URA 2464), CEA Saclay, 91191 Gif-sur-Yvette, France}

\author{Gr\'egoire de Loubens}
\thanks{Corresponding author: gregoire.deloubens@cea.fr}
\affiliation{Service de Physique de l'\'Etat Condens\'e (CNRS URA 2464), CEA Saclay, 91191 Gif-sur-Yvette, France}

\author{Olivier Klein}
\affiliation{Service de Physique de l'\'Etat Condens\'e (CNRS URA 2464), CEA Saclay, 91191 Gif-sur-Yvette, France}

\author{Andreas Riegler}
\affiliation{Physikalisches Institut (EP3), Universit\"at W\"urzburg, 97074
  W\"urzburg, Germany}

\author{Florian Lochner}
\affiliation{Physikalisches Institut (EP3), Universit\"at W\"urzburg, 97074
  W\"urzburg, Germany}

\author{Georg Schmidt}
\thanks{Present address: Institut f\"ur Physik,
  Martin-Luther-Universit\"at, Halle Wittenberg, 06099 Halle, Germany}
\affiliation{Physikalisches Institut (EP3), Universit\"at W\"urzburg, 97074
  W\"urzburg, Germany}

\author{Laurens W. Molenkamp}
\affiliation{Physikalisches Institut (EP3), Universit\"at W\"urzburg, 97074
  W\"urzburg, Germany}

\date{\today}

\begin{abstract}

  In a vortex-state magnetic
  nano-disk\cite{miramond97,cowburn99,guslienko08}, the static
  magnetization is curling in the plane, except in the core region
  where it is pointing out-of-plane\cite{shinjo00,wachowiak02}, either
  up or down leading to two possible stable states of opposite core
  polarity $p$. Dynamical reversal of $p$ by large amplitude motion of
  the vortex core has recently been demonstrated
  experimentally\cite{waeyenberge06,yamada07,curcic08,weigand09,vansteenkiste09},
  raising fundamental
  interest\cite{hertel06,hertel07,guslienko08a,lee08a} for potential
  application in magnetic storage devices\cite{pigeau10}. Here we
  demonstrate coherent control of $p$ by single and double microwave
  pulse sequences, taking advantage of the resonant vortex dynamics in
  a perpendicular bias magnetic field\cite{loubens09}.  Optimization
  of the microwave pulse duration required to switch $p$ allows to
  experimentally infer the characteristic decay time of the vortex
  core in the large oscillation regime. It is found to be more than
  twice shorter than in the small oscillation regime, raising the
  fundamental question of the non-linear behaviour of magnetic
  dissipation.

\end{abstract}

\maketitle

Magnetic vortices are topological solitons with rich dynamical
properties. The lowest energy excitation of the vortex ground state is
the so-called gyrotropic mode\cite{guslienko08}, corresponding to the
gyration of the vortex core around its equilibrium position with a
frequency in the sub-gigahertz range\cite{park03,novosad05}. It is now
established experimentally\cite{vansteenkiste09} that the excitation
of this gyrotropic motion leads to a dynamical distortion of the
vortex core profile, as predicted by micromagnetic simulations and
theoretical analysis\cite{guslienko08a}. This distortion increases
with the linear velocity of the vortex core and opposes the core
polarity, until the critical velocity $V_c\simeq 1.66\gamma
\sqrt{A_\text{ex}}$ ($\gamma$ is the gyromagnetic ratio of the
magnetic material and $A_\text{ex}$ its exchange constant) is reached
and the vortex core polarity is reversed\cite{lee08a}.

In zero magnetic field, dynamical control of the polarity is difficult
due to the degeneracy of the gyrotropic frequencies associated to
opposite polarities $p=\pm 1$, which can lead to multiple core
switching\cite{yamada07,hertel07}. Still, selective core polarity
reversal is possible using a circularly polarized microwave magnetic
field because the sense of the core rotation is linked by a right-hand
rule to its polarity\cite{curcic08}. Control of polarity switching can
also be achieved by precise timing of non resonant magnetic field
pulses\cite{weigand09,keavney09}, in a similar fashion as domain wall
propagation in magnetic nanowires\cite{thomas06}.

Resonant amplification\cite{thirion03} of the vortex gyrotropic motion
enables to reverse the core polarity with minimum excitation
power\cite{curcic08,vansteenkiste09,pigeau10}, as it allows to
concentrate the energy in a narrow frequency band. In this scheme, the
damping ratio is an important parameter because it controls the
minimum amplitude of the resonant excitation required to switch the
core\cite{lee08a}. Here, it is shown that the damping ratio close to
the reversal threshold is significantly larger than the one measured
in the small oscillation regime. We associate this to the non-linear
nature of the reversal process\cite{hertel06,guslienko08a}.

Investigation of vortex core reversal using time-resolved imaging
techniques based on X-ray magnetic circular dichroism is very
powerful, as it enables direct determination of the vortex core
trajectory\cite{vansteenkiste09}. However, it requires to average many
events and the interpretation can become more complicated due to
multiple reversal processes. In order to investigate the coupling
between resonant microwave pulses and vortex dynamics near the core
reversal threshold, we use a magnetic resonance force microscope
(MRFM) in combination with a bias magnetic field applied
perpendicularly to the disk plane, that induces two distinct resonant
gyrotropic frequencies associated to opposite core
polarities\cite{ivanov02,loubens09}. This frequency splitting is used
for a simple, single shot reading\cite{pigeau10} of the polarity state
using MRFM, before and after microwave pulses have been applied to the
vortex-state nano-disk.

The MRFM detection setup\cite{klein08}, illustrated in Fig.1a, is
described in the Methods section. It allows to probe the magnetization
dynamics in individual nanostructures\cite{loubens07}, and in
particular to spectroscopically determine the resonance frequency of
the gyrotropic mode in a vortex-state nano-disk\cite{loubens09}. The
studied nano-disk with thickness $44$~nm and diameter $1~\mu$m is made
of NiMnSb alloy (see MFM image in Fig.1b), an ultra-low damping
epitaxial material\cite{heinrich04} (typical Gilbert constant
$\alpha_{LLG}$ is between 0.002 and 0.003). A gold antenna patterned
on top allows to apply pulses of linearly polarized microwave magnetic
field $h$ in the plane of the disk.

In our experiments, we choose the bias perpendicular field $\mu_0
H=65$~mT in order to unambiguously discriminate the two different
gyrotropic frequencies, or polarity states: resonant frequencies
$f_-=217$~MHz and $f_+ =254$~MHz respectively correspond to $p=-1$
(core antiparallel to $H$, see Fig.1c) and $p=+1$ (core parallel to
$H$, see Fig.1d). The microwave power employed to read the polarity
state with MRFM is weak enough ($P=-19$~dBm), so that $p$ is not
reversed during the reading sequence\cite{pigeau10}. We also emphasize
that the bias perpendicular field $\mu_0H=65$~mT is almost five times
smaller than the static field required to switch the core
polarity\cite{thiaville03,loubens09}, so that no significant asymmetry
between the two possible reversal processes ($p=-1$ to $p=+1$ and vice
versa) is induced, as it will be clear from the data presented below.

First, we explore in Fig.2a the efficiency of single microwave pulses
to switch the vortex core, depending on their duration, frequency and
power. $\Pi_-$ pulses are defined as single pulses that reverse the
polarity from $p=-1$ to $p=+1$ and $\Pi_+$ pulses from $p=+1$ to
$p=-1$ (see Fig.2b). The experimental data is acquired as follows: for
$\Pi_-$ ($\Pi_+$) pulses, the $p=-1$ ($p=+1$) state is first reset
using an initialization pulse whose result is known to be fully
deterministic\cite{pigeau10}. Then, a single microwave pulse of given
duration $w$, frequency $f$ and power $P$ is applied, and the final
polarity state is read using MRFM.  An opaque coloured pixel marks
pulse settings for which reversals are recorded with a 100\% success
rate, while a blank pixel means that no reversal is recorded. The
contour plots presented in Fig.2a with different shades of red (blue)
show the superposition of the results for $\Pi_-$ ($\Pi_+$) pulses
with three different durations $w$: 100, 50 and 20~ns.

The frequency splitting introduced by the perpendicular bias field
clearly appears in Fig.2a.  Due to the resonant character of the
investigated switching process, a pronounced minimum in the power
level required to reverse the vortex core is observed at a frequency
$f_-^*$ ($f_+^*$) close to the gyrotropic frequency associated to the
initial core polarity, for each duration of the $\Pi_-$ ($\Pi_+$)
pulses. This frequency discrimination introduced by $H$ allows most of
the recorded results to be fully deterministic.

The minimum power $P^*$ required to reverse the core polarity
increases as the pulse duration $w$ decreases. We have plotted the
dependences on $w$ of the optimal frequency $f_+^*$ (Fig.2d) and of
the optimal pulse energy $E^*=P^*w$ (Fig.2e). In these plots, $w$
ranges from 1~$\mu$s down to 3~ns, with experimental points obtained
from the analysis of data sets similar to those presented
Fig.2a. There is a clear minimum in the optimal pulse energy that
occurs around $w=50$~ns. The position of this minimum yields an
estimation of the characteristic decay time. If the pulse duration
exceeds this characteristic time, the vortex core dynamics reaches
steady-state before the end of the pulse, i.e., the pulse duration is
uselessly too long. On the contrary, if the pulse duration is shorter
than this characteristic time, some energy is wasted outside the
resonance line.  Due to the transient response of the vortex core to
the excitation pulse, the optimal frequency $f_+^*$ measured for
pulses shorter than 50~ns shifts to higher frequency\cite{lee08a} (see
Fig.2d), a general behaviour of any harmonic oscillator forced on time
scales shorter than its decay time.

To be more quantitative, we have performed a numerical calculation
based on the analytical approach developed for zero applied magnetic
field\cite{lee08a}, with the new ingredient that due to the bias
perpendicular field, $f_-$ and $f_+$ are distinct frequencies. The
trajectory of the vortex core submitted to the microwave field pulse
is calculated using Thiele's equation\cite{thiele73}, and Guslienko's
criterion\cite{guslienko08a} for core velocity is used to determine if
the polarity is switched by the end of the pulse. The behaviour of the
optimal pulse energy as a function of $w$ can be reproduced with an
analytical formula where the damping ratio $d_\text{forced}^*$ is an
adjustable parameter (blue solid line in Fig.2e). The best fit is
obtained for $d_\text{forced}^*=0.018$, corresponding to a
characteristic decay time $\tau_\text{forced}=1/(d_\text{forced}^*2\pi
f_+)\simeq35$~ns. The predicted optimal pulse duration is
$w_\text{min}=1.26\tau_\text{forced}\simeq44$~ns (see Methods section
for details). Fig.2c illustrates the good agreement of the model with
the experimental data of Fig.2a on the whole range of pulse
parameters.

The significant result here is that the damping ratio
$d_\text{forced}^*=0.018$ inferred in the forced regime close to the
core reversal threshold is more than twice larger than the value
measured in the small oscillation regime of the gyrotropic mode, found
to be $d_\text{vortex}=0.0075$ (corresponding to a decay time
$\tau_\text{vortex}=85$~ns). The damping ratio $d_\text{vortex}$ is
itself about three times larger than the Gilbert constant, measured
from the linewidth in the perpendicularly saturated state of the
NiMnSb disk and found to be $\alpha_{LLG}=0.0025$ (see Supplementary
Figure 1). The difference between $d_\text{vortex}$ and $\alpha_{LLG}$
corresponds to the expected increase due to topological
renormalization\cite{thiele73,guslienko06}. This enhancement is
produced by the strongly inhomogeneous spatial distribution of
magnetization in the undistorted vortex state compared to the
perpendicularly saturated state. Our experimental results of Fig.2
give a quantitative estimation of the damping ratio
$d_\text{forced}^*$ for the vortex distorted by large
oscillations\cite{vansteenkiste09,guslienko08a}. In a further step, it
would be interesting to evaluate how much of the observed increase of
$d_\text{forced}^*$ against $d_\text{vortex}$ is due to topological
renormalisation between the distorted and undistorted vortex
states. As can be seen in Fig.2e, a practical consequence of this
increase is to decrease the optimal pulse duration $w_\text{min}$ and
to increase the corresponding minimum pulse energy.

Next, we investigate in Fig.3 the vortex core dynamics in the free
decay regime by monitoring the core reversal induced by two
consecutive microwave pulses. In these experiments, the pulse duration is
set to $w=9$~ns and the power to $P=-1.8$~dBm, i.e., slightly below
the minimum power $P^*=-1$~dBm required to reverse the core polarity
$p$ for such a short pulse. Contrary to a single pulse, two pulses
separated by a delay $\tau$ (Fig.3a: $\Pi_-$--$\tau$--$\Pi_-$; Fig.3b:
$\Pi_+$--$\tau$--$\Pi_+$) can reverse $p$. The striking oscillatory
dependence on the pulses carrier frequency and on the delay $\tau$
observed in Figs.3a and 3b enlightens the phase coherent coupling
between the vortex gyrotropic motion and the microwave
excitation. During the pulse duration, the gyrotropic motion is forced
at the pulse frequency, while in the free decay regime, the core
oscillates towards its equilibrium position at its natural frequency,
therefore acquiring a phase shift with respect to the excitation
carrier. As a result, the efficiency of the second pulse to drive the
vortex core to the reversal threshold, which is estimated out of ten
attempts in Fig.3, depends on the microwave frequency and on the delay
between pulses in an oscillatory manner (the oscillation period scales
as the inverse frequency detuning).

To illustrate further this effect of coherence, similar experiments
with a $\pi$ phase shift introduced between the two pulses have been
carried out. It is clear that the regions where successful reversal
are observed in Figs.3c and 3d are complementary to those in Figs.3a
and 3b, respectively. Thus, phase control of the microwave excitation
can trigger vortex core switching.

Using the same approach as before, it is possible to calculate
numerically such experimental phase diagrams. An excellent agreement
with experiments is achieved, as can be seen in Fig.4a. In Fig.4b, we
have plotted the calculated vortex core trajectory and velocity as a
function of time corresponding to two consecutive $\Pi_+$ pulses at
$f=253$~MHz separated by a 60~ns delay. In the top graphs, there is no
phase shift between pulses and the time delay is such that the second
pulse is efficient to amplify the gyrotropic motion from the beginning
of the pulse, in contrary to the bottom graphs, where a $\pi$ phase
shift is set between pulses. As a result, the vortex core is not
reversed by the end of the second pulse in the latter case, while it
is in the former. The obtained data sets also allow us to fit the free
decay time, and the best agreement with the experiment is obtained for
$\tau_\text{free} \simeq 53$~ns. Analysis of the difference of this
value with $\tau_\text{forced}$ should further shed some light on the
precise non-linear nature of magnetic dissipation close to the
reversal threshold and on its dependence on the amplitude of the
vortex core motion.

This research was partially supported by the French Grant Voice
ANR-09-NANO-006-01, EU Grants DynaMax FP6-IST-033749 and Master
NMP-FP7-212257. The authors acknowledge fruitful discussions with
V.S. Tiberkevich, A.N. Slavin and K.Y. Guslienko.

%\newpage
%%%%%%%%%%%%%%%%%%%%%%%%%%%%%%%%%%%%%%%%%%%%%%%%%%%%%%%%%%%%%%%%%%
% METHODS
%%%%%%%%%%%%%%%%%%%%%%%%%%%%%%%%%%%%%%%%%%%%%%%%%%%%%%%%%%%%%%%%%%

\section*{Methods}

  \subsection{Sample preparation.} The magnetic nano-disk (thickness
  $44$~nm, diameter $1~\mu$m) was patterned by standard e-beam
  lithography and ion-milling techniques from an extended film of
  NiMnSb grown by molecular-beam epitaxy on an InP(001)
  substrate\cite{bach03}. A 50~nm thick Si$_3$N$_4$ cap layer was
  deposited on top of the disk for protection and a broadband coplanar
  microwave antenna (300~nm thick Au) was subsequently evaporated on
  top of the patterned disk. The width of the antenna constriction
  above the disk is 5~$\mu$m (see Supplementary Figure 2).

  \subsection{Microwave setup.} Injecting a microwave current from a
  synthesizer inside the antenna produces an in-plane linearly
  polarized microwave magnetic field $h$, oriented perpendicular to
  the stripe direction. In order to apply two consecutive pulses
  separated by a variable delay $\tau$, the carrier of the cw
  excitation is split into two branches, each of them being gated by
  independent mixers before being recombined. A phase shift can be
  introduced between the two branches using a delay line. The typical
  rise and fall times of pulses are 0.8~ns and the control on the
  delay $\tau$ is better than 0.1~ns.  The calibration of the
  amplitude of $h$ yields the value $\mu_0 h=1.05$~mT with an error
  bar of 25\% for a 0~dBm input power in the antenna. Therefore, the
  microwave power range in Fig.2a corresponds to microwave field
  amplitude ranging from 0.2 to 0.7~mT, the power in double pulses
  experiments of Fig.3a to $\mu_0 h=0.85$~mT, and the power employed
  to read the polarity to $\mu_0 h=0.1$~mT.

  \subsection{Initialization pulse.} A $\Pi_+$ pulse with settings
  ($w=50$~ns, $f=f_+$, $P=-11$~dBm)\cite{pigeau10} is used to set the
  initial polarity state to $p=-1$. In fact, such an initialization
  pulse does not affect the $p=-1$ state and it transforms $p=+1$ into
  $p=-1$ (see Fig.2a). Similarly, a $\Pi_-$ pulse ($w=50$~ns, $f=f_-$,
  $P=-11$~dBm) is used to set the initial polarity state to $p=+1$.

  \subsection{MRFM detection.}
  The MRFM setup\cite{klein08} is located inside a vacuum chamber
  ($10^{-6}$~mbar) and operates at room temperature. The cantilever is
  an Olympus Biolever (spring constant $k\simeq5$~mN/m) with a 800~nm
  diameter sphere of soft amorphous Fe (with 3\% Si) glued to its
  apex. MRFM spectroscopy is achieved by placing the centre of the
  magnetic spherical probe above the centre of the NiMnSb
  nano-disk. The separation between the sample and the probe is
  $s=1.5~\mu$m. The probe senses the dipolar force $F_z$ proportional
  to the perpendicular component $M_z$ of the magnetization of the
  nano-disk (see Fig.1a). Ferromagnetic resonance spectra are obtained
  as a function of the microwave excitation frequency at a fixed bias
  field $H$. The microwave modulation is a cyclic absorption sequence,
  where the microwave power is switched on and off at the cantilever
  resonance frequency, $f_c \simeq 11.8$~kHz. The MRFM signal
  originates from the cyclic diminution of $M_z$ of the nano-disk
  synchronous with the absorption of the microwave
  field\cite{klein08,loubens09}. The resulting force modulated at the
  mechanical resonance of the cantilever force produces a cantilever
  vibration amplitude enhanced by its quality factor $Q\simeq 4000$,
  that is optically detected.

  \subsection{Numerical calculations.} The calculations presented in
  Figs.2 and 4 are based on linearised Thiele's
  equation\cite{thiele73}, an effective equation of motion for the
  vortex core position $\bm{X}$ in the disk plane:
  \begin{equation}
    -\bm{G}\times\bm{\dot{X}}-\hat{D}\bm{\dot{X}}+\kappa\bm{X}+\mu[\bm{\hat{z}}\times\bm{h}]=0.
    \label{thiele}
  \end{equation}
  The first term is the gyroforce ($\bm{G}$ is the gyrovector), the
  second one is the damping ($d=-D/|G|$), the third one is the
  restoring force ($\kappa$ is the stiffness coefficient), and the
  last one is the Zeeman energy with the spatially uniform external
  field $\bm{h}$ ($\bm{\hat{z}}$ is the unit vector normal to the disk
  plane and details on $\mu$ and previously mentioned coefficients can
  be found in refs.\cite{guslienko06,guslienko08}). The resonant
  frequency of vortex core gyration is $\omega_G=\kappa/|G|$. From
  Eq.\ref{thiele} one can find the instantaneous position $\bm{X}(t)$
  and the velocity $v(t)$ of the vortex core in the presence or
  absence of the harmonic excitation field $h(t)$, as derived in the
  supplementary documents of ref.\cite{lee08a}. If Guslienko's
  criterion on critical velocity\cite{guslienko08a} is met at some
  time of the simulated pulse sequence ($v(t)>V_c$), the vortex core
  is assumed to have switched. Experimental sample size, gyrotropic
  frequencies $f_-$ and $f_+$ associated to each polarity, and
  non-linear redshift of the frequency (3\%) are used in our
  calculation. The minimal excitation amplitude $h_c(\omega,w)$ to
  reach $V_c$ by the end of the pulse depends\cite{lee08a} on the
  excitation frequency $\omega$ and on the pulse duration
  $w$. Minimization of the energy $E \propto w h_c^2(\omega,w)$ with
  respect to $\omega$ yields the dependence on $w$ of the optimal
  frequency $f_+^*$ plotted in Fig.2d and of the optimal pulse energy
  $E^*$ plotted in Fig.2e:
  \begin{equation}
    E^*(w) = w \frac{h_c^{2}(w)}{a}=\left( \frac{6dV_c}{a\gamma R}\right)^2 \frac{w}{1+e^{-2d\omega_G w}-2e^{-d\omega_G w}}.
    \label{E*}
  \end{equation}
  In this expression, $a$ is an experimental conversion factor between
  the input power in the antenna and $h^2$ determined from
  calibration. Therefore, two independent parameters are used to
  adjust the calculation with the data in Fig.2e: the damping ratio
  $d$ to fit the overall shape (position of minimum) and the the
  critical velocity $V_c$ to fit the absolute value of the energy. The
  optimal pulse duration $w_\text{min}$ corresponding to the minimum
  energy $E^*_\text{min}$ follows from Eq.\ref{E*}:
  $w_\text{min}=1.26/(d_\text{forced}^*\omega_G)=1.26
  \tau_\text{forced}$ (for the blue solid line in Fig.2e,
  $d_\text{forced}^*=0.018$, $\tau_\text{forced}=35$~ns, hence
  $E^*_\text{min}$ is reached for $w_\text{min}=44$~ns). The fitted
  critical velocity in our NiMnSb disk is $V_c\simeq 190$~m/s, in good
  agreement with the expected value\cite{lee08a} of $225$~m/s (the
  exchange constant of NiMnSb is
  $A_\text{ex}=6~$~pJ/m\cite{ritchie03}). The same parameters are used
  in calculations of Figs.2 and 4. In Fig.4, the only new fitting
  parameter is the characteristic time $\tau_\text{free}$ of the
  vortex core free decay (the relaxation is assumed to be
  exponential). We also allow a fine adjustment of the phase shift
  between the two pulses in order to get the best agreement with the
  data (25\textdegree~in the ``zero phase shift'' experiment and
  190\textdegree~in the ``$\pi$ phase shift'' experiment; these small
  differences are ascribed to an imperfect delay line in the
  experimental pulse setup).

\newpage
%%%%%%%%%%%%%%%%%%%%%%%%%%%%%%%%%%%%%%%%%%%%%%%%%%%%%%%%%%%%%%%%%%
% REFERENCES
%%%%%%%%%%%%%%%%%%%%%%%%%%%%%%%%%%%%%%%%%%%%%%%%%%%%%%%%%%%%%%%%%%

%\newpage
%%%%%%%%%%%%%%%%%%%%%%%%%%%%%%%%%%%%%%%%%%%%%%%%%%%%%%%%%%%%%%%%%%
% FIGURES
%%%%%%%%%%%%%%%%%%%%%%%%%%%%%%%%%%%%%%%%%%%%%%%%%%%%%%%%%%%%%%%%%%

\begin{figure}
  \includegraphics[width=12cm]{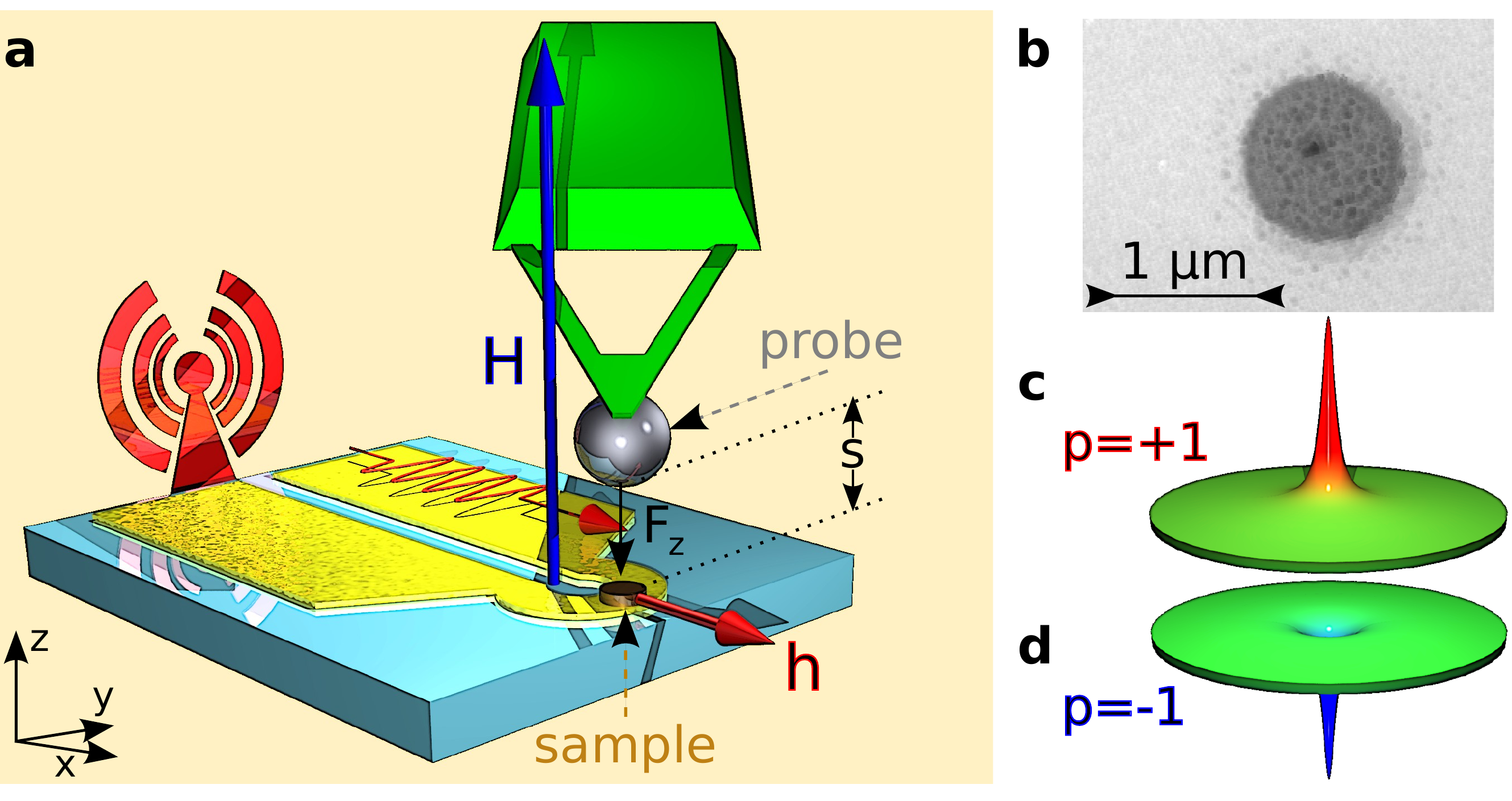}
  \caption{\small \textbf{Experimental setup and sample.} \textbf{a},
    A magnetic resonance force microscope (see Methods) is used to
    probe the vortex core dynamics of an individual vortex-state
    NiMnSb disk (diameter $1~\mu$m, thickness $44$~nm). A soft
    cantilever, with a spherical magnetic probe attached at its end
    and placed at a distance $s=1.5~\mu$m from the sample, detects
    mechanically the vortex core dynamics. The bias magnetic field $H$
    is applied perpendicularly to the disk plane whereas microwave
    pulses of linearly polarized field $h$ are applied in the
    plane. \textbf{b}, Zero-field magnetic force image of the magnetic
    vortex, where the dark spot at the disk centre reveals the core
    prepared in the $p=+1$ (\textbf{c}) polarity state. The opposite
    configuration is the $p=-1$ (\textbf{d}) polarity state.}
\end{figure}

\begin{figure}
  \includegraphics[width=12cm]{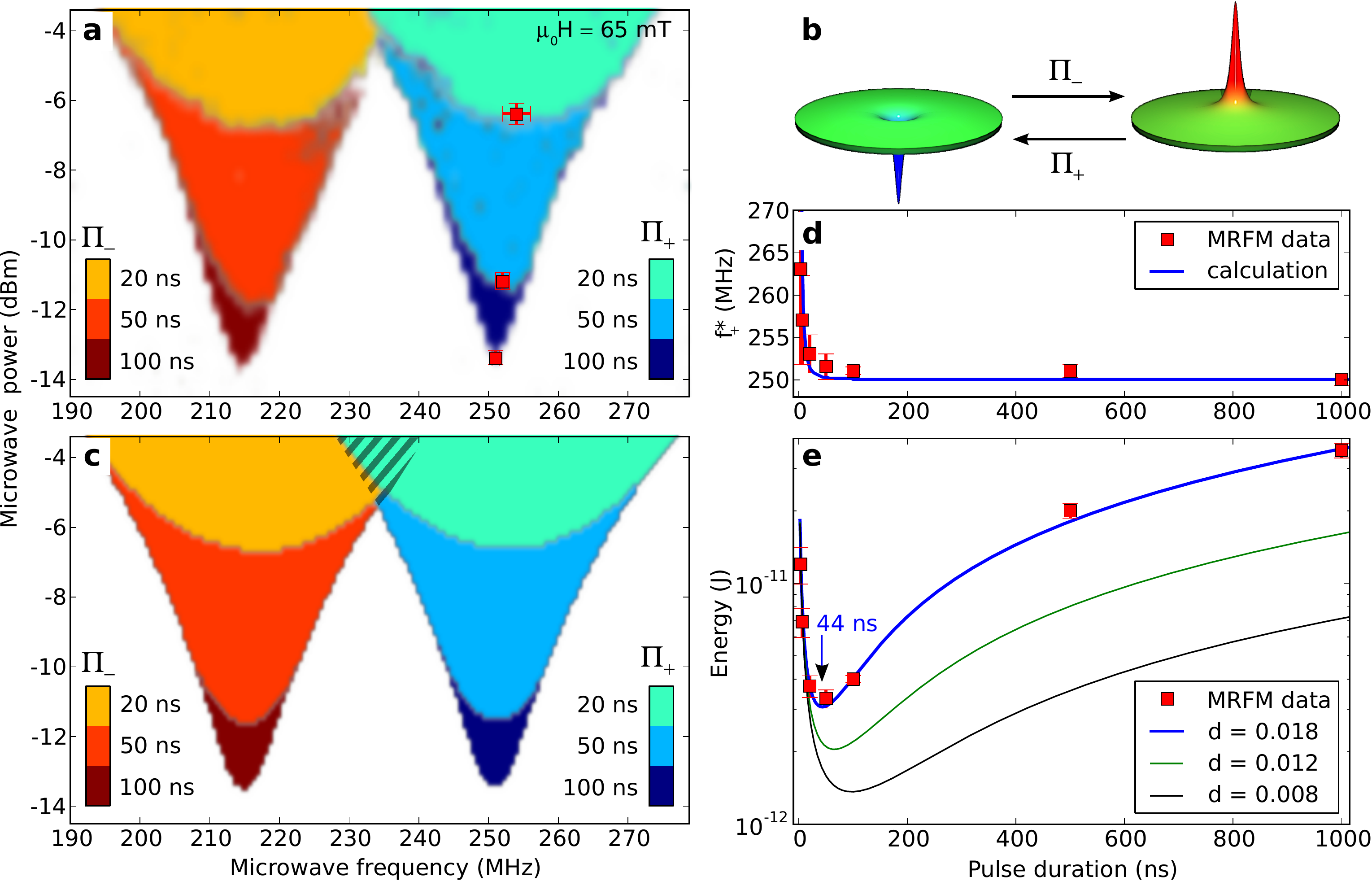}
  \caption{\small \textbf{Vortex core resonant reversal by a single
      microwave pulse.} \textbf{a}, Efficiency of the $\Pi_-$ and
    $\Pi_+$ pulses (see panel \textbf{b}) to reverse the vortex core
    at $\mu_0 H=65$~mT measured as a function of power and frequency
    (stepped by increments of 0.3~dBm and 1.2~MHz). For each pulse
    type, a coloured pixel ($\Pi_-$: red shade, $\Pi_+$: blue shade)
    marks a successful reversal. The transparency gives the switching
    probability averaged over 16 attempts. Experiments corresponding
    to three values of the pulse duration $w$ are displayed. For each
    $w$ we define ($f_+^*$,$P^*$), the optimal working point of the
    $\Pi_+$ pulse located at the bottom of the corresponding contour
    plot (see \color{red}$\blacksquare$\color{black}). \textbf{c},
    Numerical calculation (see Methods) of the experiments presented
    in \textbf{a}. This calculation is not valid within the shaded
    area, where multiple vortex core reversals can occur. \textbf{d,
      e}, Experimental (\color{red}$\blacksquare$\color{black}) and
    calculated (lines) dependences on $w$ of the optimal frequency
    $f_+^*$ (\textbf{d}) and of the optimal pulse energy $E^*=P^*w$
    (\textbf{e}). The experimental points are obtained from the
    analysis of data sets similar to those presented in \textbf{a},
    where $w$ is varied from 1~$\mu$s down to 3~ns (the three
    \color{red}$\blacksquare$\color{black}~close to the minimum energy
    in \textbf{e} are inferred from those displayed in \textbf{a}
    ). The best agreement is obtained for a damping ratio
    $d_\text{forced}^*=0.018$, i.e., a characteristic decay time
    $\tau_\text{forced}=35\pm4$~ns. The absolute value of the energy
    is also fitted in the calculation, which enables to extract the
    critical velocity for vortex core reversal, $V_c\simeq 190$~m/s
    (see Methods). Error bars on $f_+^*$ and $E^*$ are absolute minima
    and maxima resulting from experimental uncertainties in \textbf{a}
    on the optimal working point ($f_+^*$, $P^*$) associated to each
    pulse duration.}
\end{figure}

\begin{figure}
  \includegraphics[width=12cm]{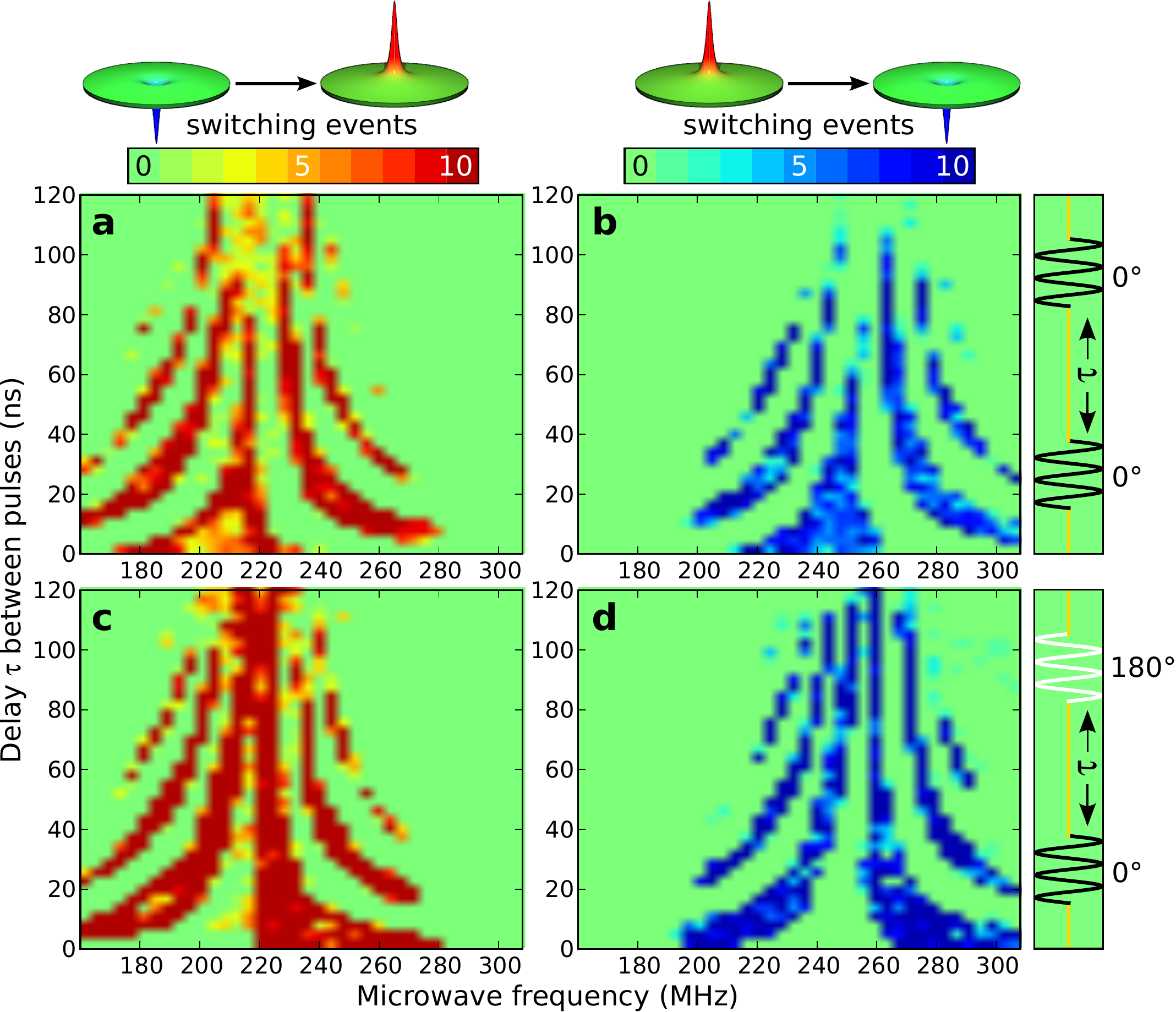}
  \caption{\small \textbf{Oscillatory dependence on frequency and
      delay between two consecutive pulses of the vortex core reversal
      efficiency.} \textbf{a--d}, Number of switching events out of
    ten attempts as a function of the delay $\tau$ separating the two
    pulses ($w=9$~ns, $P=-1.8$~dBm) and of the carrier microwave
    frequency (stepped by increments of 3~ns and 4~MHz,
    respectively). The bias magnetic field is $\mu_0 H=65$~mT. The
    initial polarity state is $p=-1$ in the left graphs (\textbf{a,
      c}) and $p=+1$ in the right graphs (\textbf{b, d}). As depicted
    in the right-side panels, the phase difference between the two
    pulses is zero in the upper graphs (\textbf{a, b}) and $\pi$ in
    the lower graphs (\textbf{c, d}).}
\end{figure}

\begin{figure}
  \includegraphics[width=16.5cm]{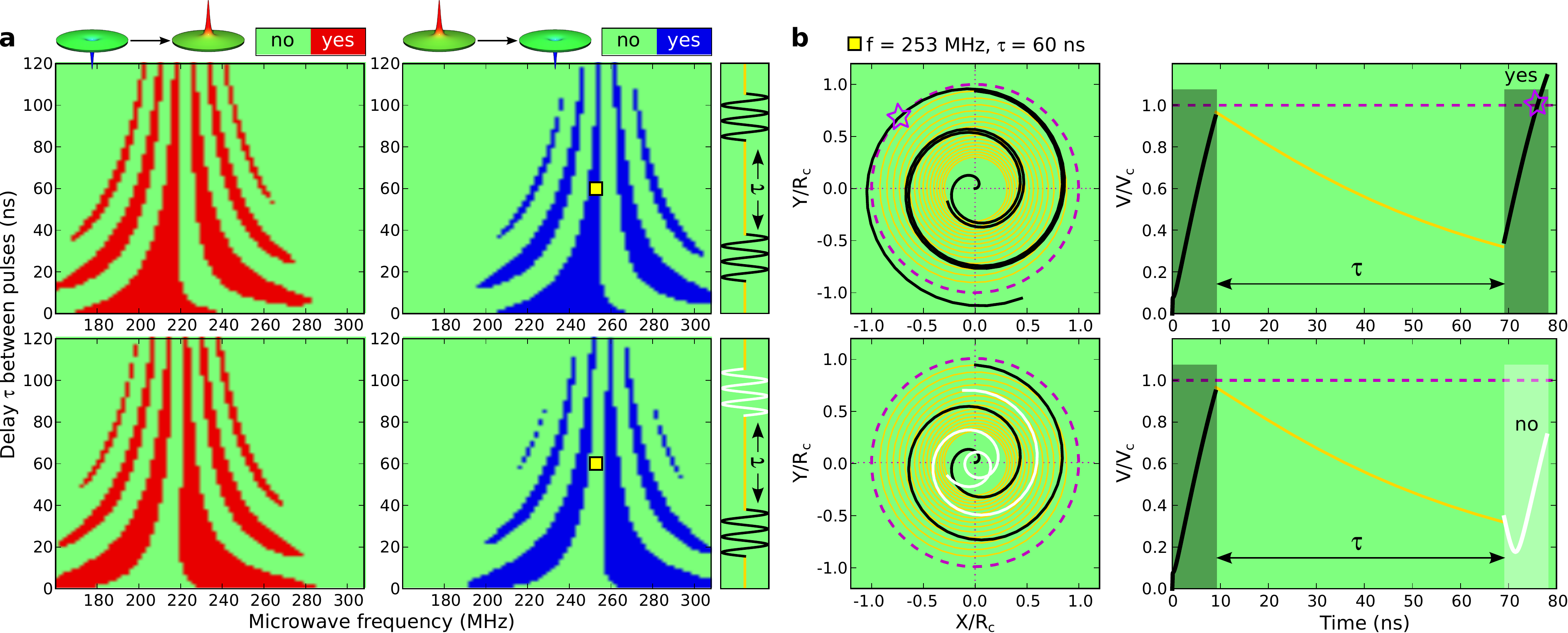}
  \caption{\small \textbf{Phase coherent control of vortex core
      reversal.}  \textbf{a}, Numerical calculation (see Methods) of
    the double pulse sequences presented in Fig.3. The best agreement
    is obtained for a characteristic decay time
    $\tau_\text{free}=53\pm6$~ns in the free regime. \textbf{b},
    Associated vortex core trajectory (left) and velocity (right)
    vs. time plotted for two $\Pi_+$ pulses with settings $\tau=60$~ns
    and $f=253$~MHz (see pixel
    \color{yellow}$\blacksquare$\color{black}~in \textbf{a}). For
    these settings, the vortex core is reversed when the phase
    difference between the pulses is equal to zero (top, see star) and
    not reversed when it is equal to $\pi$ (bottom).}
\end{figure}

\begin{figure}
  \includegraphics[width=9cm]{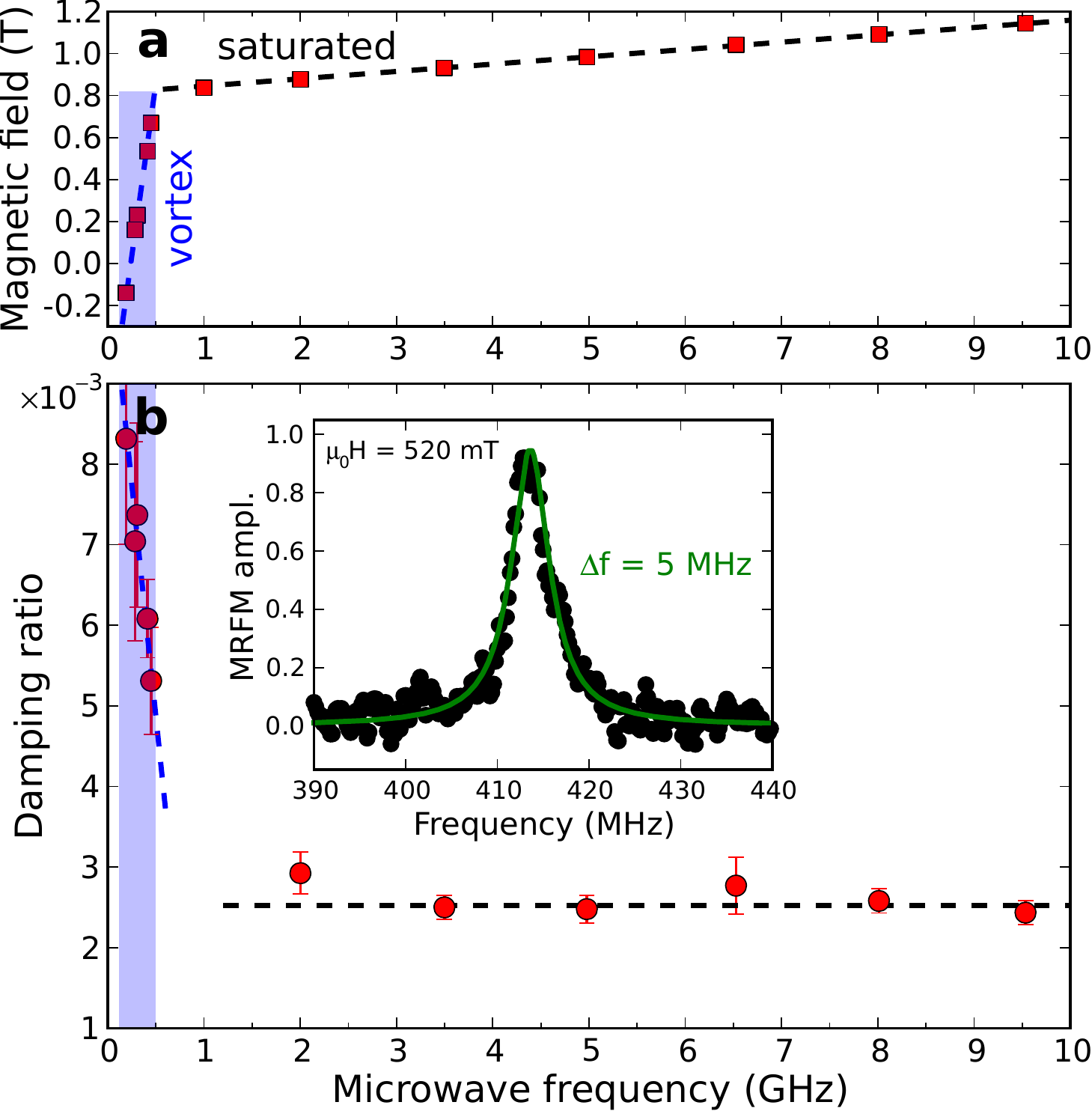}
  \caption{\small (Supplementary Figure 1) \textbf{Damping ratio in
      the small oscillation regime.}  \textbf{a},
    Frequency--perpendicular field dispersion relation of the uniform
    mode in the saturated state ($\mu_0H>0.8$~T) and of the gyrotropic
    mode in the vortex state ($\mu_0H<0.8$~T). Dashed lines are
    analytical expressions (Eqs. 1 and 2 in
    ref.\cite{loubens09}). \textbf{b}, Dependence on frequency of the
    damping ratio obtained from MRFM spectroscopic measurements in the
    small oscillation regime (the typical power level is $-30$~dBm,
    corresponding to $\mu_0h=0.03$~mT). Dashed lines are guides to the
    eye. Lorentzian fits of the resonance peaks (inset shows such a
    fit at a fixed field in the vortex state) yield the centre
    frequency $f$ and the frequency linewidth $\Delta f$ together with
    the associated error bars, hence the damping ratio $d=\Delta
    f/(2f)$.  The damping ratio in the saturated state is equal to the
    Gilbert constant $d=\alpha_{LLG}=0.0025$. In the vortex state, the
    damping ratio is renormalized by topology (Eq.13b in
    ref.\cite{thiele73}). Experimentally, it increases from about
    0.005 to 0.008 as the frequency decreases from 450~MHz down to
    190~MHz (i.e., the perpendicular field decreases from $0.65$~T
    down to $-0.15$~T). Close to $H=0$, the experimental value
    $d=d_\text{vortex}=0.0075\pm0.001$ corresponds to a decay time
    $\tau_\text{vortex}=1/(\pi \Delta f)=85\pm12$~ns and is in good
    agreement with the predicted value\cite{guslienko06}
    $d_\text{vortex}=\alpha_{LLG}\left[1+\ln{(R/b)}/2\right]\simeq0.007$
    ($R=500$~nm and $b\simeq15$~nm are respectively the disk and
    vortex core radii).}
\end{figure}

\begin{figure}
  \includegraphics[width=9cm]{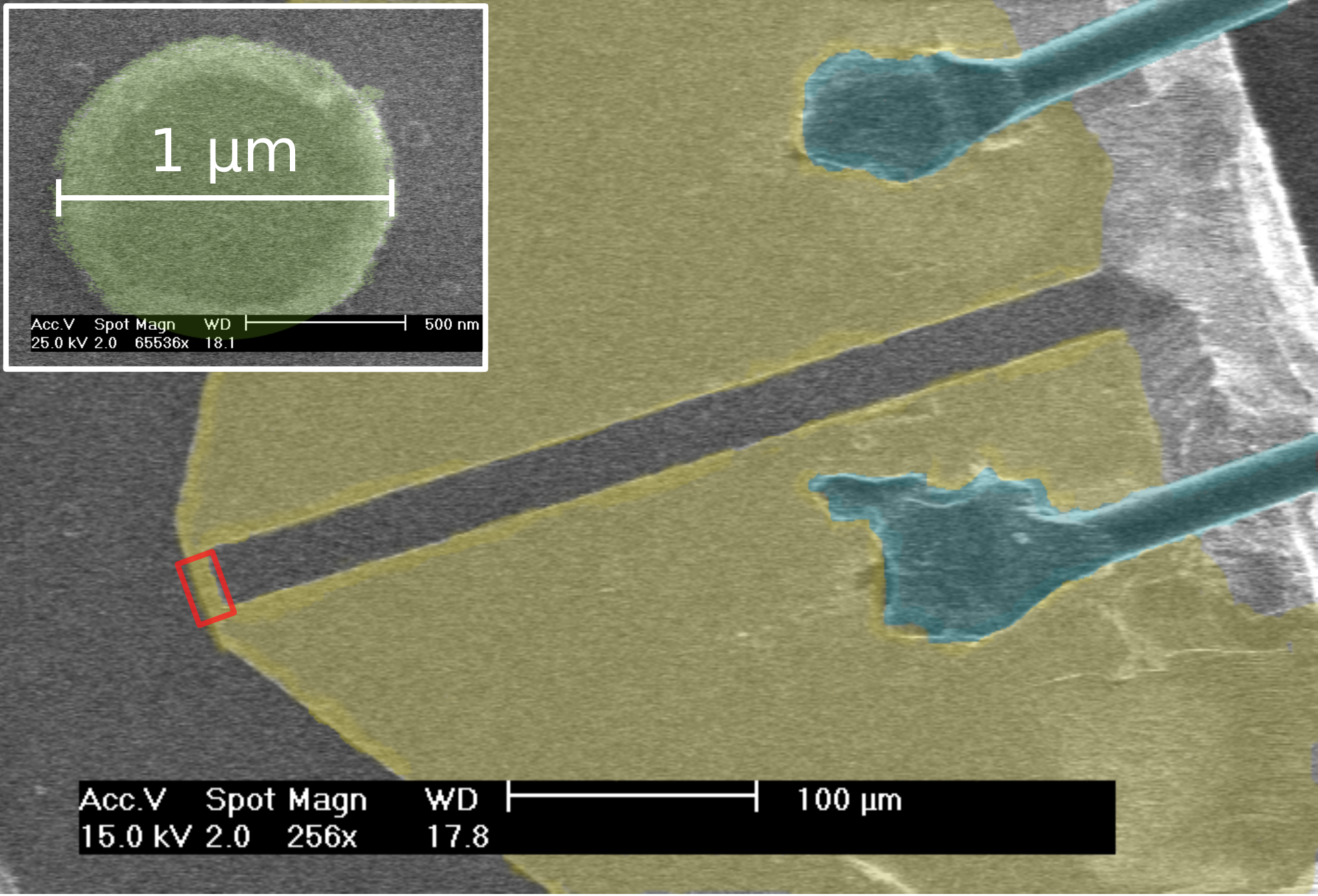}
  \caption{\small (Supplementary Figure 2) \textbf{Microwave antenna
      and NiMnSb disk sample.}  Scanning electron microscopy (SEM)
    image of the microwave antenna where the red rectangle points out
    the $5~\mu$m wide constriction under which the NiMnSb disk
    (diameter $1~\mu$m, thickness $44$~nm) is located (see
    insert). The microwave antenna is wire bounded to a coplanar
    waveguide on the right side.}
\end{figure}


%merlin.mbs 2010-03-15 4.21a (PWD, AO, DPC)
\begin{thebibliography}{0}%
\makeatletter
\providecommand \@ifxundefined [1]{%
 \@ifx{#1\undefined}
}%
\providecommand \@ifnum [1]{%
 \ifnum #1\expandafter \@firstoftwo
 \else \expandafter \@secondoftwo
 \fi
}%
\providecommand \@ifx [1]{%
 \ifx #1\expandafter \@firstoftwo
 \else \expandafter \@secondoftwo
 \fi
}%
\providecommand \natexlab [1]{#1}%
\providecommand \enquote  [1]{``#1''}%
\providecommand \bibnamefont  [1]{#1}%
\providecommand \bibfnamefont [1]{#1}%
\providecommand \citenamefont [1]{#1}%
\providecommand \href@noop [0]{\@secondoftwo}%
\providecommand \href [0]{\begingroup \@sanitize@url \@href}%
\providecommand \@href[1]{\@@startlink{#1}\@@href}%
\providecommand \@@href[1]{\endgroup#1\@@endlink}%
\providecommand \@sanitize@url [0]{\catcode `\\12\catcode `\$12\catcode
  `\&12\catcode `\#12\catcode `\^12\catcode `\_12\catcode `\%12\relax}%
\providecommand \@@startlink[1]{}%
\providecommand \@@endlink[0]{}%
\providecommand \url  [0]{\begingroup\@sanitize@url \@url }%
\providecommand \@url [1]{\endgroup\@href {#1}{\urlprefix }}%
\providecommand \urlprefix  [0]{URL }%
\providecommand \Eprint [0]{\href }%
\@ifxundefined \urlstyle {%
  \providecommand \doi  [0]{\begingroup \@sanitize@url \@doi}%
  \providecommand \@doi [1]{\endgroup \@@startlink {\doibase
  #1}doi:\discretionary {}{}{}#1\@@endlink }%
}{%
  \providecommand \doi  [0]{doi:\discretionary{}{}{}\begingroup
  \urlstyle{rm}\Url }%
}%
\providecommand \doibase [0]{http://dx.doi.org/}%
\providecommand \Doi [0]{\begingroup \@sanitize@url \@Doi }%
\providecommand \@Doi  [1]{\endgroup\@@startlink{\doibase#1}\@@Doi}%
\providecommand \@@Doi [1]{#1\@@endlink}%
\providecommand \selectlanguage [0]{\@gobble}%
\providecommand \bibinfo  [0]{\@secondoftwo}%
\providecommand \bibfield  [0]{\@secondoftwo}%
\providecommand \translation [1]{[#1]}%
\providecommand \BibitemOpen [0]{}%
\providecommand \bibitemStop [0]{}%
\providecommand \bibitemNoStop [0]{.\EOS\space}%
\providecommand \EOS [0]{\spacefactor3000\relax}%
\providecommand \BibitemShut  [1]{\csname bibitem#1\endcsname}%
%</preamble>
\end{thebibliography}%


\begin{thebibliography}{10}
\expandafter\ifx\csname url\endcsname\relax
  \def\url#1{\texttt{#1}}\fi
\expandafter\ifx\csname urlprefix\endcsname\relax\def\urlprefix{URL }\fi
\providecommand{\bibinfo}[2]{#2}
\providecommand{\eprint}[2][]{\url{#2}}

\bibitem{miramond97}
\bibinfo{author}{Miramond, C.}, \bibinfo{author}{Fermon, C.},
  \bibinfo{author}{Rousseaux, F.}, \bibinfo{author}{Decanini, D.} \&
  \bibinfo{author}{Carcenac, F.}
\newblock \bibinfo{title}{Permalloy cylindrical submicron size dot arrays}.
\newblock \emph{\bibinfo{journal}{J. Magn. Magn. Mater.}}
  \textbf{\bibinfo{volume}{165}}, \bibinfo{pages}{500--503}
  (\bibinfo{year}{1997}).

\bibitem{cowburn99}
\bibinfo{author}{Cowburn, R.~P.}, \bibinfo{author}{Koltsov, D.~K.},
  \bibinfo{author}{Adeyeye, A.~O.}, \bibinfo{author}{Welland, M.~E.} \&
  \bibinfo{author}{Tricker, D.~M.}
\newblock \bibinfo{title}{Single-domain circular nanomagnets}.
\newblock \emph{\bibinfo{journal}{Phys. Rev. Lett.}}
  \textbf{\bibinfo{volume}{83}}, \bibinfo{pages}{1042--1045}
  (\bibinfo{year}{1999}).

\bibitem{guslienko08}
\bibinfo{author}{Guslienko, K.~Y.}
\newblock \bibinfo{title}{Magnetic vortex state stability, reversal and
  dynamics in restricted geometries}.
\newblock \emph{\bibinfo{journal}{J. Nanosci. Nanotechnol.}}
  \textbf{\bibinfo{volume}{8}}, \bibinfo{pages}{2745--2760}
  (\bibinfo{year}{2008}).

\bibitem{shinjo00}
\bibinfo{author}{Shinjo, T.}, \bibinfo{author}{Okuno, T.},
  \bibinfo{author}{Hassdorf, R.}, \bibinfo{author}{Shigeto, K.} \&
  \bibinfo{author}{Ono, T.}
\newblock \bibinfo{title}{Magnetic vortex core observation in circular dots of
  permalloy}.
\newblock \emph{\bibinfo{journal}{Science}} \textbf{\bibinfo{volume}{289}},
  \bibinfo{pages}{930--932} (\bibinfo{year}{2000}).

\bibitem{wachowiak02}
\bibinfo{author}{Wachowiak, A.} \emph{et~al.}
\newblock \bibinfo{title}{Direct observation of internal spin structure of
  magnetic vortex cores}.
\newblock \emph{\bibinfo{journal}{Science}} \textbf{\bibinfo{volume}{298}},
  \bibinfo{pages}{577--580} (\bibinfo{year}{2002}).

\bibitem{waeyenberge06}
\bibinfo{author}{Waeyenberge, B.~V.} \emph{et~al.}
\newblock \bibinfo{title}{Magnetic vortex core reversal by excitation with
  short bursts of an alternating field}.
\newblock \emph{\bibinfo{journal}{Nature}} \textbf{\bibinfo{volume}{444}},
  \bibinfo{pages}{461--464} (\bibinfo{year}{2006}).

\bibitem{yamada07}
\bibinfo{author}{Yamada, K.} \emph{et~al.}
\newblock \bibinfo{title}{Electrical switching of the vortex core in a magnetic
  disk}.
\newblock \emph{\bibinfo{journal}{Nature Mater.}} \textbf{\bibinfo{volume}{6}},
  \bibinfo{pages}{270--273} (\bibinfo{year}{2007}).

\bibitem{curcic08}
\bibinfo{author}{Curcic, M.} \emph{et~al.}
\newblock \bibinfo{title}{Polarization selective magnetic vortex dynamics and
  core reversal in rotating magnetic fields}.
\newblock \emph{\bibinfo{journal}{Phys. Rev. Lett.}}
  \textbf{\bibinfo{volume}{101}}, \bibinfo{pages}{197204}
  (\bibinfo{year}{2008}).

\bibitem{weigand09}
\bibinfo{author}{Weigand, M.} \emph{et~al.}
\newblock \bibinfo{title}{Vortex core switching by coherent excitation with
  single in-plane magnetic field pulses}.
\newblock \emph{\bibinfo{journal}{Phys. Rev. Lett.}}
  \textbf{\bibinfo{volume}{102}}, \bibinfo{pages}{077201}
  (\bibinfo{year}{2009}).

\bibitem{vansteenkiste09}
\bibinfo{author}{Vansteenkiste, A.} \emph{et~al.}
\newblock \bibinfo{title}{X-ray imaging of the dynamic magnetic vortex core
  deformation}.
\newblock \emph{\bibinfo{journal}{Nature Physics}}
  \textbf{\bibinfo{volume}{5}}, \bibinfo{pages}{332--334}
  (\bibinfo{year}{2009}).

\bibitem{hertel06}
\bibinfo{author}{Hertel, R.} \& \bibinfo{author}{Schneider, C.~M.}
\newblock \bibinfo{title}{Exchange explosions: Magnetization dynamics during
  vortex-antivortex annihilation}.
\newblock \emph{\bibinfo{journal}{Phys. Rev. Lett.}}
  \textbf{\bibinfo{volume}{97}}, \bibinfo{pages}{177202}
  (\bibinfo{year}{2006}).

\bibitem{hertel07}
\bibinfo{author}{Hertel, R.}, \bibinfo{author}{Gliga, S.},
  \bibinfo{author}{F\"{a}hnle, M.} \& \bibinfo{author}{Schneider, C.~M.}
\newblock \bibinfo{title}{Ultrafast nanomagnetic toggle switching of vortex
  cores}.
\newblock \emph{\bibinfo{journal}{Phys. Rev. Lett.}}
  \textbf{\bibinfo{volume}{98}}, \bibinfo{pages}{117201}
  (\bibinfo{year}{2007}).

\bibitem{guslienko08a}
\bibinfo{author}{Guslienko, K.~Y.}, \bibinfo{author}{Lee, K.-S.} \&
  \bibinfo{author}{Kim, S.-K.}
\newblock \bibinfo{title}{Dynamic origin of vortex core switching in soft
  magnetic nanodots}.
\newblock \emph{\bibinfo{journal}{Phys. Rev. Lett.}}
  \textbf{\bibinfo{volume}{100}}, \bibinfo{pages}{027203}
  (\bibinfo{year}{2008}).

\bibitem{lee08a}
\bibinfo{author}{Lee, K.-S.} \emph{et~al.}
\newblock \bibinfo{title}{Universal criterion and phase diagram for switching a
  magnetic vortex core in soft magnetic nanodots}.
\newblock \emph{\bibinfo{journal}{Phys. Rev. Lett.}}
  \textbf{\bibinfo{volume}{101}}, \bibinfo{pages}{267206}
  (\bibinfo{year}{2008}).

\bibitem{pigeau10}
\bibinfo{author}{Pigeau, B.} \emph{et~al.}
\newblock \bibinfo{title}{A frequency-controlled magnetic vortex memory}.
\newblock \emph{\bibinfo{journal}{Appl. Phys. Lett.}}
  \textbf{\bibinfo{volume}{96}}, \bibinfo{pages}{132506}
  (\bibinfo{year}{2010}).

\bibitem{loubens09}
\bibinfo{author}{de~Loubens, G.} \emph{et~al.}
\newblock \bibinfo{title}{Bistability of vortex core dynamics in a single
  perpendicularly magnetized nanodisk}.
\newblock \emph{\bibinfo{journal}{Phys. Rev. Lett.}}
  \textbf{\bibinfo{volume}{102}}, \bibinfo{pages}{177602}
  (\bibinfo{year}{2009}).

\bibitem{park03}
\bibinfo{author}{Park, J.~P.}, \bibinfo{author}{Eames, P.},
  \bibinfo{author}{Engebretson, D.~M.}, \bibinfo{author}{Berezovsky, J.} \&
  \bibinfo{author}{Crowell, P.}
\newblock \bibinfo{title}{Imaging of spin dynamics in closure domain and vortex
  structure}.
\newblock \emph{\bibinfo{journal}{Phys. Rev. B}} \textbf{\bibinfo{volume}{67}},
  \bibinfo{pages}{020403(R)} (\bibinfo{year}{2003}).

\bibitem{novosad05}
\bibinfo{author}{Novosad, V.} \emph{et~al.}
\newblock \bibinfo{title}{Magnetic vortex resonance in patterned ferromagnetic
  dots}.
\newblock \emph{\bibinfo{journal}{Phys. Rev. B}} \textbf{\bibinfo{volume}{72}},
  \bibinfo{pages}{024455} (\bibinfo{year}{2005}).

\bibitem{keavney09}
\bibinfo{author}{Keavney, D.~J.}, \bibinfo{author}{Cheng, X.~M.} \&
  \bibinfo{author}{Buchanan, K.~S.}
\newblock \bibinfo{title}{Polarity reversal of a magnetic vortex core by a
  unipolar, nonresonant in-plane pulsed magnetic field}.
\newblock \emph{\bibinfo{journal}{Appl. Phys. Lett.}}
  \textbf{\bibinfo{volume}{94}}, \bibinfo{pages}{172506}
  (\bibinfo{year}{2009}).

\bibitem{thomas06}
\bibinfo{author}{Thomas, L.} \emph{et~al.}
\newblock \bibinfo{title}{Oscillatory dependence of current-driven magnetic
  domain wall motion on current pulse length}.
\newblock \emph{\bibinfo{journal}{Nature}} \textbf{\bibinfo{volume}{443}},
  \bibinfo{pages}{197--200} (\bibinfo{year}{2006}).

\bibitem{thirion03}
\bibinfo{author}{Thirion, C.}, \bibinfo{author}{Wernsdorfer, W.} \&
  \bibinfo{author}{Mailly, D.}
\newblock \bibinfo{title}{Switching of magnetization by nonlinear resonance
  studied in single nanoparticles}.
\newblock \emph{\bibinfo{journal}{Nature Mater.}} \textbf{\bibinfo{volume}{2}},
  \bibinfo{pages}{524--527} (\bibinfo{year}{2003}).

\bibitem{ivanov02}
\bibinfo{author}{Ivanov, B.~A.} \& \bibinfo{author}{Wysin, G.~M.}
\newblock \bibinfo{title}{Magnon modes for a circular two-dimensional
  easy-plane ferromagnet in the cone state}.
\newblock \emph{\bibinfo{journal}{Phys. Rev. B}} \textbf{\bibinfo{volume}{65}},
  \bibinfo{pages}{134434} (\bibinfo{year}{2002}).

\bibitem{klein08}
\bibinfo{author}{Klein, O.} \emph{et~al.}
\newblock \bibinfo{title}{Ferromagnetic resonance force spectroscopy of
  individual submicron-size samples}.
\newblock \emph{\bibinfo{journal}{Phys. Rev. B}} \textbf{\bibinfo{volume}{78}},
  \bibinfo{pages}{144410} (\bibinfo{year}{2008}).

\bibitem{loubens07}
\bibinfo{author}{de~Loubens, G.} \emph{et~al.}
\newblock \bibinfo{title}{Magnetic {R}esonance {S}tudies of the {F}undamental
  {S}pin-{W}ave {M}odes in {I}ndividual {S}ubmicron {C}u/{N}i{F}e/{C}u
  {P}erpendicularly {M}agnetized {D}isks}.
\newblock \emph{\bibinfo{journal}{Phys. Rev. Lett.}}
  \textbf{\bibinfo{volume}{98}}, \bibinfo{pages}{127601}
  (\bibinfo{year}{2007}).

\bibitem{heinrich04}
\bibinfo{author}{Heinrich, B.} \emph{et~al.}
\newblock \bibinfo{title}{Magnetic properties of {NiMnSb}(001) films grown on
  {InGaAs/InP}(001)}.
\newblock \emph{\bibinfo{journal}{J. Appl. Phys.}}
  \textbf{\bibinfo{volume}{95}}, \bibinfo{pages}{7462} (\bibinfo{year}{2004}).

\bibitem{thiaville03}
\bibinfo{author}{Thiaville, A.}, \bibinfo{author}{Garc\'ia, J.~M.},
  \bibinfo{author}{Dittrich, R.}, \bibinfo{author}{Miltat, J.} \&
  \bibinfo{author}{Schrefl, T.}
\newblock \bibinfo{title}{Micromagnetic study of bloch-point-mediated vortex
  core reversal}.
\newblock \emph{\bibinfo{journal}{Phys. Rev. B}} \textbf{\bibinfo{volume}{67}},
  \bibinfo{pages}{094410} (\bibinfo{year}{2003}).

\bibitem{thiele73}
\bibinfo{author}{Thiele, A.~A.}
\newblock \bibinfo{title}{Steady-state motion of magnetic domains}.
\newblock \emph{\bibinfo{journal}{Phys. Rev. Lett.}}
  \textbf{\bibinfo{volume}{30}}, \bibinfo{pages}{230--233}
  (\bibinfo{year}{1973}).

\bibitem{guslienko06}
\bibinfo{author}{Guslienko, K.~Y.}
\newblock \bibinfo{title}{Low-frequency vortex dynamic susceptibility and
  relaxation in mesoscopic ferromagnetic dots}.
\newblock \emph{\bibinfo{journal}{Appl. Phys. Lett.}}
  \textbf{\bibinfo{volume}{89}}, \bibinfo{pages}{022510}
  (\bibinfo{year}{2006}).

\bibitem{bach03}
\bibinfo{author}{Bach, P.} \emph{et~al.}
\newblock \bibinfo{title}{Molecular-beam epitaxy of the half-heusler alloy
  {NiMnSb} on {(In,Ga)As/InP(001)}}.
\newblock \emph{\bibinfo{journal}{Appl. Phys. Lett.}}
  \textbf{\bibinfo{volume}{83}}, \bibinfo{pages}{521} (\bibinfo{year}{2003}).

\bibitem{ritchie03}
\bibinfo{author}{Ritchie, L.} \emph{et~al.}
\newblock \bibinfo{title}{Magnetic, structural, and transport properties of the
  heusler alloys {Co2MnSi} and {NiMnSb}}.
\newblock \emph{\bibinfo{journal}{Phys. Rev. B}} \textbf{\bibinfo{volume}{68}},
  \bibinfo{pages}{104430} (\bibinfo{year}{2003}).

\end{thebibliography}
\end{document}